\documentclass[reprint,aps]{revtex4-2}

\usepackage[T1]{fontenc}
\usepackage[utf8]{inputenc}
\usepackage{amsmath,amssymb,bm}
\usepackage{graphicx}
\usepackage{hyperref}
\usepackage{dblfloatfix}
\usepackage{float} % <-- force figures with [H]

\begin{document}

\title{Variational quantum eigensolver for chemical molecules}

\author{Luca Ion}
\affiliation{School of Physics \& Astronomy, University of Nottingham, Nottingham, UK}

\author{Adam Smith}
\affiliation{School of Physics \& Astronomy, University of Nottingham, Nottingham, UK}

\date{May 16, 2023}

\begin{abstract}
Solving multi-particle systems is an important topic in quantum chemistry and condensed matter
physics. In this article, we focus on finding the ground states and ground-state energies for the He-H$^{+}$
and H$_2$O molecules using quantum computing. We employ the variational quantum
eigensolver (VQE), which is executed both on a quantum computer simulator and on an
IBM quantum device. We compare the results against the exact ground-state energy obtained
through classical methods. The H$_2$O simulations were run on Nottingham’s High Performance
Computer (HPC).
\end{abstract}

\maketitle

\section{INTRODUCTION}
The second quantum revolution~\cite{ref1} gave rise to the
fields of quantum information and quantum computation. At present, we find ourselves in the era of
Noisy Intermediate-Scale Quantum (NISQ)~\cite{ref2} devices,
meaning that quantum devices are built from a few
hundred qubits and suffer from qubit errors (noise). Nevertheless, this era has provided new technologies
and disciplines such as quantum cryptography~\cite{ref3} and quantum machine learning~\cite{ref4}.

An important field that benefits from quantum computation is the quantum chemistry of molecules. The reason
is that the Hilbert space grows exponentially with
the number of qubits $N$, and therefore complex molecules
comprising many particles require substantial resources to store and manipulate on a classical computer
using classical methods. In contrast, quantum computation allows us to work with quantum circuits and quantum
gates, which can simplify the process.

In this work, we focus on the VQE algorithm used to find the ground
state and ground-state energy of a multi-particle Hamiltonian. We first consider the He-H$^{+}$ molecule, which
was studied extensively in~\cite{ref5}. We introduce the circuit
ansatz and discuss how the ansatz differs when a deeper
circuit structure is desired. We then discuss the results
of VQE optimization. Finally, we study the more complex H$_2$O molecule to assess how VQE performs for a larger
system.

\section{EXACT DIAGONALIZATION}
First, to find the ground-state energy for the Hamiltonian associated with the molecule under inspection,
one can use an eigenvalue decomposition. The
method used to obtain the Hamiltonians for various
molecules is discussed in Sec.~III. Once the Hamiltonian
$H$ is known, one can use classical methods to find
the ground-state energy. We used \texttt{NumPy} for
all mathematical operations, including eigenvalue decomposition and the computation of the exact ground-state energy.
\newline
\section{FINDING THE HAMILTONIAN}
The first step is to obtain the Hamiltonian for the molecule
under inspection. For He-H$^{+}$, we use the Hamiltonian studied in
\cite{ref5}; this is a two-qubit Hamiltonian and is
shown in Sec.~VI. For more complex molecules,
such as H$_2$O, we use the built-in PennyLane \texttt{molecularHamiltonian} method~\cite{ref6}.

\section{QUANTUM CIRCUIT ANSATZ}
\begin{figure}[H]
  \centering
  \includegraphics[width=0.95\textwidth]{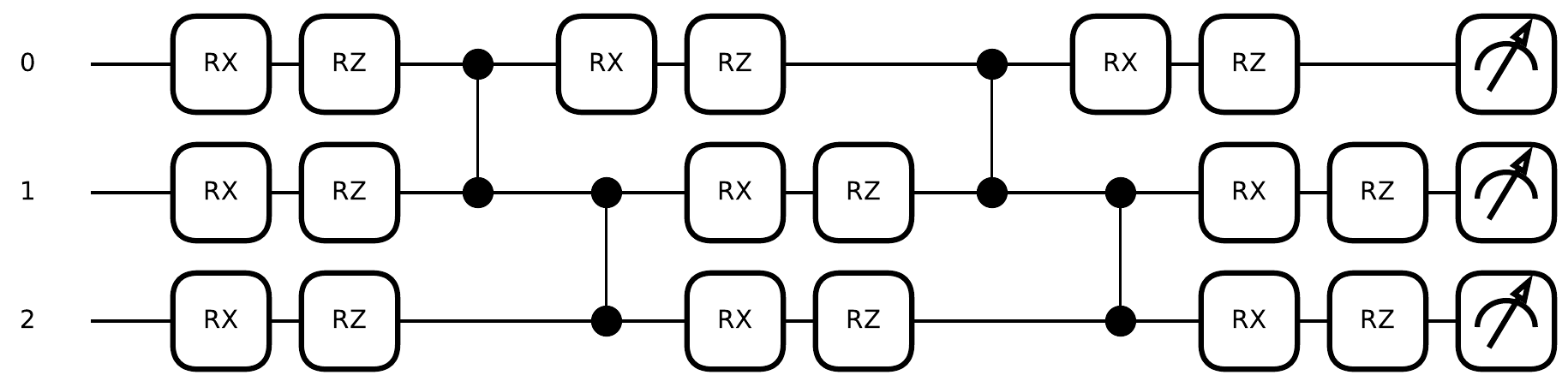}
  \caption{Circuit ansatz for $N=3$ and $M=2$. $R_X$ and $R_Z$ are rotation gates and the connection between the wires is done through Control-Z gates.}
  \label{fig:ansatz}
\end{figure}

The main goal is to carry out optimization on a quantum circuit to obtain a state that approximates the ground
state of the Hamiltonian in question. The first task is to
construct a quantum circuit structure that is able to
approximate the ground state. The circuit structure used
in this paper can be extended over multiple layers
$M$ and multiple qubits $N$. The circuit in Fig.~\ref{fig:ansatz}
is an example of such a structure. The symbols $R_X$ and
$R_Z$ denote single-qubit rotation gates,
\begin{align}
R_X(\theta_a) &= \exp\left(-i\frac{\theta_a}{2}X\right), \label{eq:rx}\\
R_Z(\theta_b) &= \exp\!\left(-i\frac{\theta_b}{2}Z\right), \label{eq:rz}
\end{align}
where $X$ and $Z$ are Pauli operators and $\theta_a$, $\theta_b$ are
rotation angles. Note that each rotation gate takes an
angle as an argument. The full set of such angles
constitutes the parameters to be optimized so that the
output state approximates the ground state.

In addition, we require entangling operations to make individual
qubits ``communicate''; this is achieved using the
$CZ$ (Control-Z) gate, a two-qubit gate with the first qubit as the control
and the second as the target. In matrix form,
\begin{equation}
CZ =
\begin{pmatrix}
1 & 0 & 0 & 0\\
0 & 1 & 0 & 0\\
0 & 0 & 1 & 0\\
0 & 0 & 0 & -1
\end{pmatrix}.
\label{eq:cz}
\end{equation}
The $CZ$ gate applies the $Z$ operation to the target qubit if
the control qubit is $\lvert 1\rangle$, and acts trivially otherwise. In
Fig.~\ref{fig:ansatz}, all connections between wires are $CZ$ gates.

\onecolumngrid
\twocolumngrid

\section{METHOD}
In order to approximate the ground-state energy, we employ the VQE
algorithm. The ansatz circuit is optimized such that the state produced by the circuit $\lvert\psi\rangle$
minimizes the energy
\begin{equation}
E=\langle \psi|H|\psi\rangle.
\label{eq:energy}
\end{equation}
It is important to note that $\lvert\psi\rangle$ and, as a result, $E$
depend on the parameter list $\vec{\theta}$; in machine learning language, $E$ is the loss function. The optimization method used
here is gradient descent~\cite{ref7} and variations of it. For classical gradient descent,
the parameter list is updated over multiple iterations according to
\begin{equation}
\vec{\theta}_{n+1}=\vec{\theta}_n-\eta\nabla E(\vec{\theta}_n),
\label{eq:gd}
\end{equation}
where $\eta$ is the learning rate. To approximate
the gradient, multiple methods can be used. In this paper, we compare the following gradient descent approximation schemes: first
order finite difference (FOGD)~\cite{ref8}, second order finite difference (SOGD)~\cite{ref8}, simultaneous perturbation stochastic approximation (SPSA)~\cite{ref9}, and parameter shift (PS)~\cite{ref10}.

Choosing a tuning method for $\eta$ is not an easy task;
here a constant $\eta$ is used for FOGD, SOGD, and PS, and a decaying $\eta$ is used for SPSA. Other tuning methods may
be used~\cite{ref11}. The number of parameters to optimize depends on the number of qubits $N$ and the number of
layers in the circuit ansatz $M$ according to
\begin{equation}
\textit{parameters} = 2N(M+1).
\label{eq:params}
\end{equation}
Parameters are initialized randomly, with each parameter initialized to $2\pi$ times a
random number between 0 and 1. The number of optimization iterations is fixed, although a stopping criterion could also be implemented.
Finally, the number of circuit evaluations
(shots) is set to 8192.

\section{HE-H$^{+}$ RESULTS}
The first molecule considered is He-H$^{+}$ (helium hydride). This molecule’s Hamiltonian depends on the distance between the atoms $R$. The Hamiltonian is a two-qubit Hamiltonian given by
\begin{equation}
\begin{aligned}
H=\frac{1}{2}[
&J_x(X1+1X)+J_z(Z1+1Z)\\
&+J_{xx}XX+J_{zz}ZZ+J_{xz}(XZ+ZX)+C]
\end{aligned}
\label{eq:heh}
\end{equation}
where $J_x$, $J_z$, $J_{xx}$, $J_{zz}$, and $C$ are parameters dependent on $R$.

\begin{figure}[H]
  \centering
  \includegraphics[width=\columnwidth]{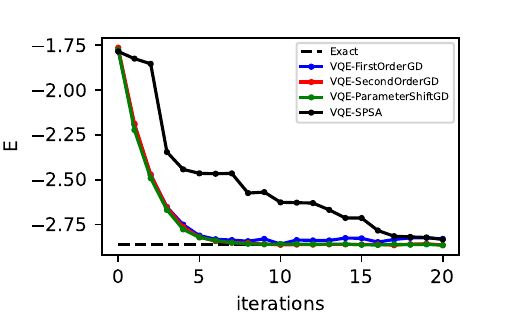}
  \caption{Simulated He-H$^{+}$ optimization for $R=0.9$\AA, $M=1$, $N=2$, $\eta=0.8$ and for all the different gradient descent variants.}
  \label{fig:heh_sim}
\end{figure}

For this molecule, we use $\eta=0.8$ and 20 iterations in
the optimization. The optimization plots for each method are shown in Fig.~\ref{fig:heh_sim}. The best
convergence is observed for the parameter-shift and SOGD methods. SPSA decreases rapidly in the first few iterations but then
stagnates. These observations depend on the tuning of $\eta$, and selecting an appropriate learning rate is therefore an important task.

Additionally, optimization can be carried out for different ansatz depths $M$ and the final energies compared. For this molecule, $M=1$ is sufficient for convergence, so increasing $M$ does not yield substantially better approximations.

\begin{figure}[H]
  \centering
  \includegraphics[width=\columnwidth]{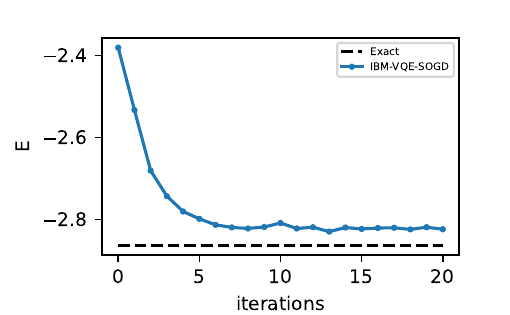}
  \caption{IBM He-H$^{+}$ optimization for $R=0.9$\AA, $M=1$, $N=2$, $\eta=0.8$ and SOGD.}
  \label{fig:heh_ibm}
\end{figure}

To run the optimization on a real IBM machine, it is impractical to evaluate all methods due to long queue times; therefore, SOGD was used since it performed well in simulation. Figure~\ref{fig:heh_ibm} shows the optimization results. The final energy does not match the exact energy closely, which reflects the noisy nature of current quantum hardware.

\begin{figure}[H]
  \centering
  \includegraphics[width=\columnwidth]{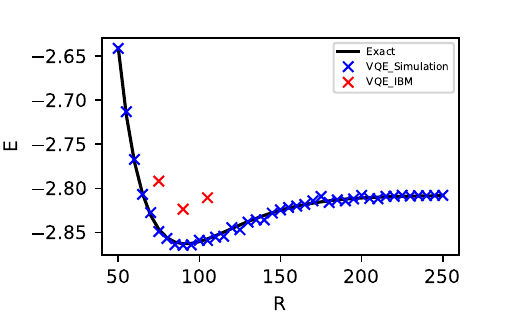}
  \caption{He-H$^{+}$: Energy($E$) against $R$ in picometres plot for the exact, VQE-simulation (SOGD), and IBM results using $M=1$ ansatz and $\eta=0.8$.}
  \label{fig:heh_curve}
\end{figure}

The optimization was also carried out for $R=0.75$\AA\ and $R=1.05$\AA\ on the real IBM machine and was run in simulation for every $R\in[0.5,2.5]$ with step size 0.05. In Fig.~\ref{fig:heh_curve}, the approximated energy is plotted for each $R$ and compared against the exact energy obtained through exact diagonalization. The energy curve has a minimum at $R=0.9$\AA, which motivates our choice of this value for the other plots in this section.

\section{H$_2$O RESULTS}
The next molecule considered is water, H$_2$O. We consider a simplified model~\cite{ref12} where the
active orbitals and active electrons are both set to 4.
With these constraints, PennyLane returns an eight-qubit
Hamiltonian for this system; without limiting the active orbitals, a fourteen-qubit Hamiltonian is obtained. Since this
system is significantly larger than the He-H$^{+}$ system, the optimization takes considerably longer to complete,
even for a $M=1$ ansatz. As a result, the HPC was used to benefit from parallelism.

\begin{figure}[H]
  \centering
  \includegraphics[width=0.65\columnwidth]{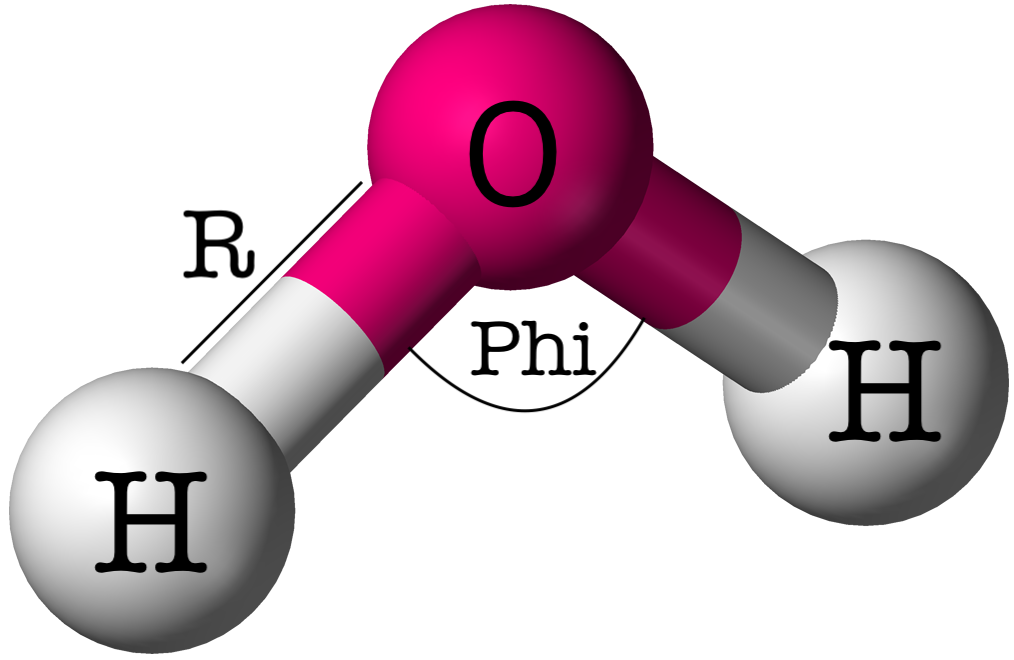}
  \caption{The geometry of the H$_2$O molecule. $R$ is the O-H bond length and $\phi$ is the HOH angle.}
  \label{fig:h2o_geom}
\end{figure}

The water molecule requires two parameters to describe its geometry: $R$ and $\phi$, where $R$ is the O--H bond length and $\phi$
is the H--O--H angle, as shown in Fig.~\ref{fig:h2o_geom}. Different values of these parameters produce different Hamiltonians and thus different ground-state energies. When examining the Hamiltonians returned by PennyLane, we found that the lowest ground-state energy occurs for $R=1.9$\AA\ and $\phi=1.75$ rad. This differs from typical literature values ($R=0.96$\AA\ and $\phi=1.8$ rad) due to the approximations introduced by limiting the active orbitals and active electrons.

\begin{figure}[H]
  \centering
  \includegraphics[width=\columnwidth]{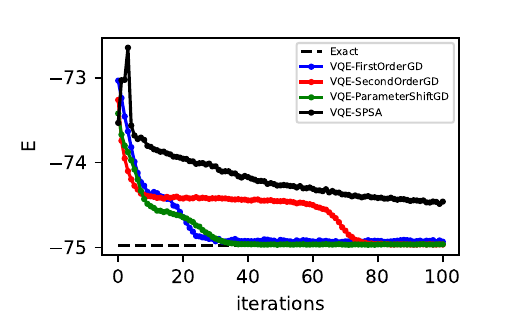}
  \caption{Simulated H$_2$O optimization for $\phi=1.75$ rad, $R=1.9$\AA, $M=1$, $N=8$, $\eta=0.8$ and for all the different gradient descent variants.}
  \label{fig:h2o_opt}
\end{figure}

Optimization was run for $M=1,2,3,4,5,8$ layered
ansatz circuits for 100 iterations with $\eta=0.8$. It was generally
found that even a $M=1$ ansatz produces good approximations. Figure~\ref{fig:h2o_opt} shows the optimization for each method.
All methods converge successfully except SPSA, which likely requires improved tuning beyond a simple decay schedule.

\begin{figure}[H]
  \centering
  \includegraphics[width=\columnwidth]{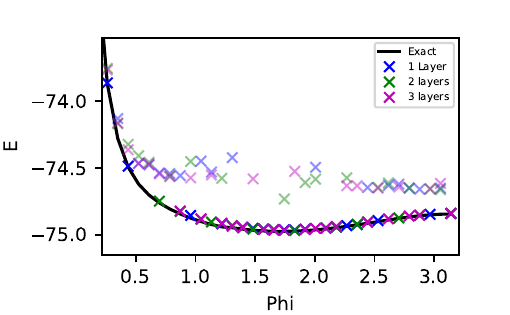}
  \caption{H$_2$O: Energy($E$) against $\phi$ plot for the exact and VQE-simulation (PS) results using $M=1,2,3$ layered ansatz and $\eta=0.8$. The more transparent data corresponds to instances that did not converge in the optimization.}
  \label{fig:h2o_phi}
\end{figure}

To show the dependence of the ground-state energy on the angle $\phi$, the bond length is fixed to
$R=1.9$\AA\ and the angle is varied from $5^{\circ}$ to $180^{\circ}$ in $5^{\circ}$ increments.
Optimization is then run for each angle and both
the exact energy and the approximated energy are plotted.
By running optimization multiple times, it was observed
that the PS method generally performed best; therefore, it is used here. Figure~\ref{fig:h2o_phi} shows the results for
$M=1,2,3$ ansatz circuits. A dip in the energy at $\phi=1.75$ rad is observed, analogous to the minimum seen for He-H$^{+}$ in Fig.~\ref{fig:heh_curve}. For some values of $\phi$, optimization becomes trapped and does not converge within the allotted number of sweeps. Addressing this likely requires improved learning-rate tuning and longer, repeated optimizations.

\section{DISCUSSION}
In this work, we carried out VQE for He-H$^{+}$ and
H$_2$O using different ansatz circuits and multiple gradient descent methods. We find that VQE can successfully approximate the ground-state energy even for a complex molecule such as H$_2$O. For the simpler He-H$^{+}$ molecule,
the algorithm was also performed on a real IBM machine,
highlighting the noisy nature of current quantum devices in the optimization results.

From the optimization plots, we observe that PS generally performs best, while SPSA performs worst under the settings considered. PS has the advantage of computing an exact gradient, whereas SPSA has an intrinsically stochastic character. On the other hand, SPSA can be faster in terms of required circuit calls, as it uses considerably fewer circuit evaluations than the other methods. For He-H$^{+}$ this speed-up is not pronounced due to the small circuit size, but for H$_2$O the difference is substantial. Moreover, SPSA is often favoured on real IBM devices because each circuit evaluation incurs queueing time; fewer circuit calls can therefore reduce total wall-clock time. However, SPSA introduces additional hyperparameters, and the challenge of tuning them is correspondingly greater.

Further investigation is needed for the H$_2$O molecule. First, improved hyperparameter tuning and repeated optimization runs may increase convergence across more of the points in Fig.~\ref{fig:h2o_phi}. Another modification is to change the initialization strategy for the ansatz. Instead of fully random rotations, one can initialize to small rotations $\vec{\theta}_0 \approx \vec{0}$, so that the initial output state is close to a computational basis state and therefore exhibits low entanglement. This is motivated by the possibility that the true ground state also has relatively low entanglement, making convergence easier. Next, it would be interesting to keep $\phi$ fixed and vary $R$ to obtain the energy as a function of bond length. Finally, increasing the active-orbital space to obtain a more realistic H$_2$O model (which will increase $N$) and performing optimization for that larger instance would be a natural next step.

\section{acknowledgments and comments}
I would like to thank my supervisor for their guidance and support throughout this mini-project, as well as the University of Nottingham for providing access to high-performance computing (HPC) resources. This work was carried out as part of my MSc degree and is intended as a learning exercise rather than a contribution of novel research.

\end{document}